\documentclass[aps,prx,reprint,superscriptaddress,longbibliography]{revtex4-2}

\usepackage{amsmath,amssymb}
\usepackage{graphicx}
\usepackage{bm}
\usepackage{hyperref}
\hypersetup{hidelinks,pdfkeywords={proximity effects, spin-orbit coupling, Wannier functions, downfolding, graphene heterostructures}}

\begin{document}

\title{First-Principles Wannier Representation of Proximity Effects}

\author{Yaroslav Zhumagulov}
\email{yaroslav.zhumagulov@epfl.ch}
\affiliation{Institute of Physics, \'Ecole Polytechnique F\'ed\'erale de Lausanne (EPFL), CH-1015 Lausanne, Switzerland}

\author{Johan F\'elisaz}
\affiliation{Institute of Physics, \'Ecole Polytechnique F\'ed\'erale de Lausanne (EPFL), CH-1015 Lausanne, Switzerland}

\author{Stepan S. Tsirkin}
\affiliation{Institute of Physics, \'Ecole Polytechnique F\'ed\'erale de Lausanne (EPFL), CH-1015 Lausanne, Switzerland}

\author{Denis Kochan}
\affiliation{Department of Physics and Center for Quantum Frontiers of Research and Technology (QFort), National Cheng Kung University, Tainan 70101, Taiwan}
\affiliation{Institute of Physics, Slovak Academy of Sciences, 84511 Bratislava, Slovakia}

\author{Oleg V. Yazyev}
\affiliation{Institute of Physics, \'Ecole Polytechnique F\'ed\'erale de Lausanne (EPFL), CH-1015 Lausanne, Switzerland}

\date{\today}

\begin{abstract}
Proximity effects in layered heterostructures are commonly represented by static parameters fitted to first-principles band structures, an approximation that usually discards the energy dependence of interlayer hybridization, momentum transfer, and spatial structure.
We address this limitation by deriving a dynamical proximity operator directly from density functional theory by downfolding the heterostructure Kohn-Sham Hamiltonian onto a fixed low-energy target Wannier subspace.
We demonstrate the method for three different graphene heterostructures coupled to magnetic and semiconducting substrates.
Graphene/hBN/Co(0001) exhibits proximity-induced exchange resonances driven by hybridization with Co states. 
Graphene/PtSe$_2$ develops a sublattice-selective, intervalley-coherent, proximity-induced coupling.
For graphene/WSe$_2$, we identify the spatial modulation of both Rashba and valley-Zeeman proximity-induced spin-orbit couplings.
The method determines the full microscopic structure of the proximity coupling from first principles, providing input for transport and correlated-phase calculations.
\end{abstract}

\maketitle

\section{Introduction}
Stacking crystals into van der Waals heterostructures turns the interface into a design parameter. Interlayer hybridization modifies spin-orbit coupling, induces exchange fields in nonmagnetic layers, and breaks target-layer sublattice symmetries; the resulting interactions depend on the material combination, stacking order, and twist angle~\cite{Geim2013, Novoselov2016, Zutic2019}.
Graphene is an extreme case: its intrinsic spin-orbit coupling (SOC) is on the $\mu$eV scale~\cite{Konschuh2010, Han2014}, and proximity can enhance it by orders of magnitude.
Transition-metal dichalcogenide (TMDC) substrates can raise the effective SOC in graphene to the meV scale~\cite{Avsar2014, Wang2015, Gmitra2016}, and magnetic substrates and hBN/ferromagnet tunnel barriers can induce a strong resonant exchange field in graphene~\cite{Yang2013, Zollner2016, Wang2015AHE, Avsar2011SpinValve, Kamalakar2014HBNTunnel, Ringer2018CoMgO}.
Consequently, the calculation of spin transport and correlated phases in proximitized graphene heterostructures requires a parametrization of the proximity-induced couplings~\cite{Gmitra2017BLGWSe2, Zollner2021BLGEncapsulated, Zollner2022TLG, Zhumagulov2024RTG, Zhumagulov2024BBG, Seiler2025LayerSelective, Zhou2021HalfQuarterMetals, Zhou2021RTGSuperconductivity, Yang2025RhombohedralSOC, Patterson2025SpinCanting, Choi2025RhombohedralQAHSC}.

First-principles calculations can describe the electronic structure of the entire heterostructure. However, proximity-induced couplings are commonly extracted by fitting a symmetry-inspired effective low-energy model to the calculated bands~\cite{Gmitra2013, Gmitra2016QSH, Zollner2016, Kochan2017Model, Naimer2021, Zollner2023, MendietaAlvarez2026HelicalSpintronics}.
Such fitting procedures can yield compact and informative parametrizations of Rashba, valley-Zeeman, staggered, and exchange couplings.
The predefined operator basis and the energy-independent form of the fitted coupling can leave three features of the microscopic structure unresolved.
First, interlayer hybridization with substrate states can generate resonant energy dependence, particularly at metallic or strongly hybridized interfaces~\cite{Wehling2010Resonant, Kochan2014}.
Second, lattice mismatch or relative twist enables finite momentum transfer and can couple valleys when the interface periodicity permits it~\cite{BistritzerMacDonald2011, AndreiMacDonald2020}.
Finally, variations in local stacking spatially modulate the proximity coupling across the interface. 

Existing first-principles scattering methods~\cite{Lu2019, Lu2020, Guo2025}, inspired by Wannier interpolation of electron-phonon coupling~\cite{Giustino2017}, retain the momentum and spatial structure of the proximity-coupling matrix elements. These methods include only the direct projection of the heterostructure Kohn--Sham Hamiltonian onto a target Wannier subspace and omit interlayer hybridization contributions. In the proximity problem, substrate states are excluded; as a result, the portion of the interaction arising from hybridization with substrate states, along with its energy dependence, is lost.

Here we include the hybridization contribution by downfolding the heterostructure Kohn--Sham Hamiltonian onto the fixed low-energy Wannier basis of the isolated target layer, without band fitting or re-wannierizing the heterostructure.
The resulting dynamical proximity operator incorporates both direct and hybridization contributions to the proximity-induced coupling matrix elements, preserving the full momentum, spatial, and energy structure.
We evaluate the hybridization term using a multishift Krylov method on a Matsubara-frequency grid and project the operator onto spin, sublattice, and valley operators.

We examine three features of the proximity coupling in graphene: resonant exchange in graphene on hBN/Co(0001), intervalley-coherent coupling in graphene on PtSe$_2$, and spatially varying Rashba and valley-Zeeman couplings in graphene on WSe$_2$. Graphene's small intrinsic SOC lets proximity-induced spin couplings dominate, while its distinct $\mathbf K$ and $\mathbf K'$ valleys make intervalley scattering well-defined.

\section{Method}
\label{sec:method}

We obtain the Kohn--Sham Hamiltonians $H_0$ and $H$ from density functional theory (DFT) calculations for the isolated target layer in its primitive cell and the heterostructure in a commensurate supercell, respectively. Both calculations use the same cutoffs and norm-conserving pseudopotentials~\cite{Pseudodojo2018}; Sec.~S1 of the Supplemental Material~\cite{SM} gives the remaining computational details~\cite{Giannozzi2009, Giannozzi2017, Perdew1996PBE, Methfessel1989, Pizzi2020, Grimme2010, Grimme2011}. We replicate the target-layer Kohn--Sham potential onto the supercell and align $H_0$ and $H$ at a common Fermi level. The proximity perturbation $V=H-H_0$ then contains the differences in the self-consistent local Kohn--Sham potentials and in the nonlocal pseudopotential operators, including their SOC contributions.

The low-energy subspace is spanned by the target maximally localized Wannier functions (MLWFs)~\cite{Marzari2012}. Their Bloch-gauge states $|w_{n\mathbf{k}}\rangle$ define the projector $P=\sum_{n,\mathbf{k}}|w_{n\mathbf{k}}\rangle\langle w_{n\mathbf{k}}|$ and its complement $Q=1-P$. For graphene, $P$ projects onto the carbon $p_z$ manifold, the standard Wannier subspace for its Dirac bands.
A properly disentangled subspace is invariant under $H_0$, so $V$ alone couples $P$ and $Q$, and L\"owdin partitioning~\cite{Lowdin1951} gives the energy-dependent proximity operator
\begin{equation}
 \mathcal{V}(z) = \underbrace{PVP}_{\text{direct}} + \underbrace{PVQ\frac{1}{z-QHQ}QVP}_{\text{interlayer hybridization}},
 \label{eq:Vop}
\end{equation}
with $z$ a complex frequency.
The direct term is the projection of the self-consistent perturbation onto $P$. The interlayer-hybridization term couples $P$ to the states of $Q$ through the projected resolvent $(z-QHQ)^{-1}$, which resums this hybridization to all orders in $V$.
We evaluate both terms in the symmetry-adapted MLWF basis of the graphene layer~\cite{Tsirkin2021}, so that expanding $\mathcal V(z)$ in symmetry-resolved operators [Eq.~\eqref{eq:sm-projection} below] identifies the proximity-induced symmetry-breaking couplings.

Rather than inverting $z-QHQ$, we obtain the hybridization term from a shifted resolvent~\cite{Baroni2001} at the fermionic Matsubara frequencies $z=i\omega_\ell$, with energies measured from the Fermi level $E_F$. Here $\omega_\ell=(2\ell+1)\pi/\beta$, where $\beta=(k_{\rm B}T)^{-1}$ is the inverse temperature and $k_{\rm B}$ is the Boltzmann constant.
For each state $|w_{n\mathbf k_2}\rangle$ we solve the coupled linear system
\begin{equation}
\begin{split}
 &\sum_{\mathbf k_3}\left[H_{\mathbf k_1\mathbf k_3}+(\alpha P-i\omega_\ell)\delta_{\mathbf k_1\mathbf k_3}\right]|\Delta\psi_{n\mathbf k_2}(\mathbf k_3,i\omega_\ell)\rangle\\
 &\qquad=-QV_{\mathbf k_1\mathbf k_2}|w_{n\mathbf k_2}\rangle,
\end{split}
\label{eq:resolvent}
\end{equation}
where $\mathbf k_1$ and $\mathbf k_3$ run over the momenta coupled to $\mathbf k_2$. Supercell periodicity allows $H$ and $V$ to fold primitive-cell momenta when $\mathbf k_1-\mathbf k_2=\mathbf G_s$, where $\mathbf G_s$ is a reciprocal-lattice vector of the supercell, whereas the projectors $P$ and $Q$ are momentum-diagonal. The solution $|\Delta\psi_{n\mathbf k_2}(\mathbf k_3,i\omega_\ell)\rangle$ is the response at $\mathbf k_3$ generated by $|w_{n\mathbf k_2}\rangle$.
The energy shift $\alpha>0$ suppresses its component in $P$, and its $Q$ component approaches $(z-QHQ)^{-1}QV|w_{n\mathbf k_2}\rangle$ as $\alpha$ increases, so the finite-shift solve approximates the projected resolvent with an error of order $\alpha^{-1}$~\cite{SM}.
The downfolded matrix elements then read
\begin{equation}
\begin{split}
 &\mathcal V_{mn}(\mathbf k_1,\mathbf k_2;i\omega_\ell)=\langle w_{m\mathbf k_1}|V|w_{n\mathbf k_2}\rangle\\
 &\quad+\sum_{\mathbf k_3}\langle w_{m\mathbf k_1}|VQ|\Delta\psi_{n\mathbf k_2}(\mathbf k_3,i\omega_\ell)\rangle.
\end{split}
\label{eq:Vmatrix}
\end{equation}
Elements with $\mathbf k_1\ne\mathbf k_2$ describe coherent scattering between primitive-cell Bloch states that fold onto the same supercell momentum.

Equation~\eqref{eq:resolvent} has to be solved for every source state and every Matsubara frequency. We write this family of linear systems as $(A+\sigma_\ell\mathbf 1)|\Delta\psi_\ell\rangle=b$, with $A=H+\alpha P$, $\sigma_\ell=-i\omega_\ell$, and $b=-QV|w_{n\mathbf k}\rangle$. A scalar shift leaves the Krylov subspace unchanged,
\begin{equation}
  \mathcal{K}_m(A + \sigma_\ell\mathbf{1}, b) = \mathcal{K}_m(A, b),
\end{equation}
where $\mathcal K_m(A,b)=\operatorname{span}\{b,Ab,\dots,A^{m-1}b\}$ is the Krylov subspace of dimension $m$, so a single multishift GMRES iteration~\cite{Frommer1998} generates one Krylov basis for all sampled frequencies: the products of the Hamiltonian with the basis vectors are shared, and only the shifted solution vectors differ.

The sampled frequencies form the sparse grid of the fermionic kernel's intermediate representation (IR)~\cite{Shinaoka2017}. At fixed truncation accuracy, the number $L$ of retained singular functions grows logarithmically with $\beta\Lambda$, where $\Lambda$ is the energy cutoff of the spectral interval. We expand each element of the hybridization correction $\mathcal V^{\rm cor}(z)=\mathcal V(z)-PVP$ in the IR basis functions $u_l$,
\begin{equation}
  \mathcal{V}^{\rm cor}_{mn}(i\omega_\ell)
  = \sum_l c_{mn,l}\,u_l(i\omega_\ell),
\end{equation}
and determine the coefficients $c_{mn,l}$ from its values at the sparse frequencies selected by the IR sampling rule~\cite{Li2020,Wallerberger2023}.
Because the coefficients fix $\mathcal V^{\rm cor}$ at every Matsubara frequency, the sampled values carry its full frequency dependence at the truncation accuracy.
For each Wannier matrix element, a Pad\'e approximant constructed from the values at the positive sampled frequencies and evaluated at $\omega+i0^+$ gives the real-frequency coupling~\cite{Vidberg1977}. The Matsubara values carry no stochastic noise, which is the main source of instability in Pad\'e continuation, and the agreement of the continued spectra with the DFT bands [Figs.~\ref{fig:hbnco}(c) and \ref{fig:ptse2}(b)] checks the continuation.

To connect $\mathcal V$ with effective low-energy models, we project a proximity matrix $X(z)$ onto orthogonal Hermitian operators $\mathcal O_a$. The coefficient of $\mathcal O_a$ is
\begin{equation}
 c_a(z)=\frac{\operatorname{Tr}[\mathcal O_a^\dagger X(z)]}
 {\operatorname{Tr}[\mathcal O_a^\dagger\mathcal O_a]},
 \label{eq:sm-projection}
\end{equation}
where the trace runs over the chosen low-energy basis and $X$ and $\mathcal O_a$ are expressed in the same Wannier gauge.
Projecting $\mathcal V$ yields the SOC and exchange coefficients, while the spin-independent reference $H_0$ contributes to the dispersion and the Dirac-point offset; the momentum derivatives of $H_0$ at the valley centers determine the graphene Fermi velocity.
The spectral functions are calculated from the full matrix $\mathcal V(\omega)$, not from its projected parts. Sections~S2 and S3 of the Supplemental Material~\cite{SM} give the convergence tests and the individual channel projections.

\section{Results}
\label{sec:results}

\subsection{Graphene on hBN/Co(0001): resonant proximity exchange coupling}

\begin{figure*}[t]
\includegraphics[width=\textwidth]{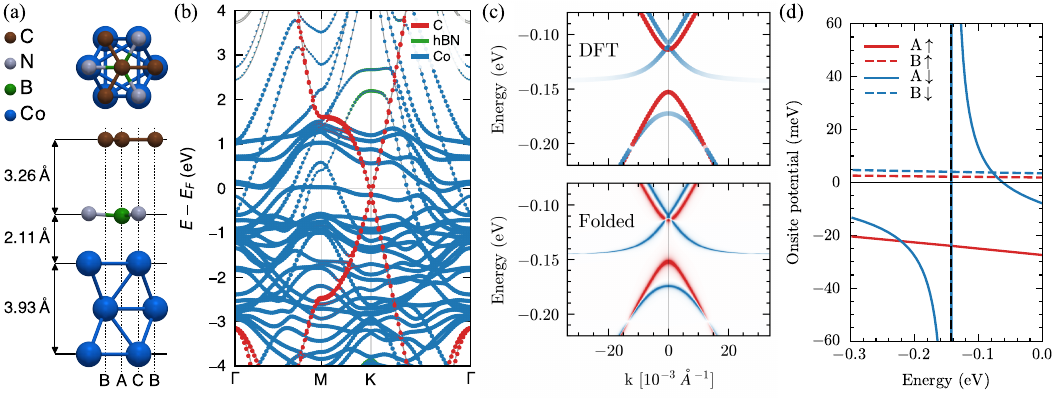}
  \caption{
    Graphene/hBN/Co(0001) proximity results. (a)~Atomic structure: graphene on a single hBN layer on an hcp Co(0001) surface, modeled by three Co layers. Arrows give the distances between the layer mean planes; the dashed lines mark the three stacking columns of the hexagonal surface: $A$ carries the graphene $A$ sublattice above the boron, $B$ the graphene $B$ sublattice above the second Co layer, and $C$ the nitrogen and the outer Co layers.
    (b)~Layer-projected band structure of the heterostructure along $\Gamma{-}M{-}K{-}\Gamma$; color distinguishes graphene from hBN and Co states.
    (c)~Zoom on the Dirac point: DFT bands of the heterostructure (upper) and the spectrum calculated from the downfolded Hamiltonian with broadening $\eta = 2$~meV (lower), color-coded by the spin projection $\langle s_z\rangle$ of the graphene states.
    (d)~Real part of the diagonal elements of the hybridization correction $\mathcal{V}^{\rm cor}(\mathbf{K},\mathbf{K};\omega)$ for the four graphene Wannier states $\{A{\uparrow},\,B{\uparrow},\,A{\downarrow},\,B{\downarrow}\}$ versus real frequency; the frequency-independent direct term is not included. Vertical dashed lines mark substrate-dominated Kohn--Sham levels associated with the Co $d$ resonances.
  }
  \label{fig:hbnco}
\end{figure*}
Graphene on hBN on an hcp Co(0001) surface exhibits a resonant proximity-induced exchange coupling in its Dirac bands~\cite{Zollner2016, Zollner2025}.
Such stacks are used as spin-injection contacts in graphene devices, where the tunnel barrier improves the injection efficiency while the ferromagnetic metal induces local proximity fields~\cite{Avsar2011SpinValve, Kamalakar2014HBNTunnel, Gurram2017, Ringer2018CoMgO}.
The exchange splitting arises from hybridization between Co $d$ orbitals and graphene $p_z$ states and varies with the number of hBN layers and their stacking order~\cite{Zollner2016}.
The lattice constants of graphene, hBN, and the hcp Co(0001) basal plane are nearly commensurate, so the heterostructure fits within a shared hexagonal unit cell~\cite{SM}.
As a consequence, $\mathcal{V}_{mn}(\mathbf{k}_1,\mathbf{k}_2;\omega)$ is diagonal in crystal momentum.
We consider crystal momenta $\mathbf k$ in the vicinity of $\mathbf K$, where the intravalley proximity coupling varies little with momentum over the region relevant to the Dirac bands. We therefore retain its frequency dependence and approximate $\mathcal{V}(\mathbf{k},\mathbf{k};\omega) \approx \mathcal{V}(\mathbf{K},\mathbf{K};\omega)$~\cite{Gmitra2016, Zollner2016}, giving the Green function
\begin{equation}
  G^{-1}(\mathbf{k},\omega) = \omega+i\eta - H_0(\mathbf{k}) - \mathcal{V}(\mathbf{K},\mathbf{K};\omega),
\label{eq:G-eff}
\end{equation}
with $H_0(\mathbf{k})$ the graphene Wannier Hamiltonian referenced to the Fermi level in the spin-sublattice basis
$\{A{\uparrow},\,B{\uparrow},\,A{\downarrow},\,B{\downarrow}\}$, and $\eta$ a broadening.

The layer-projected bands in Fig.~\ref{fig:hbnco}(b) show the graphene Dirac cone embedded in the Co $d$ and hBN manifolds with which it hybridizes.
We resolve the spin character using the spectral function $A_s(\mathbf{k},\omega)=-\pi^{-1}\operatorname{Im}\operatorname{Tr}[s_z\,\sigma_0G(\mathbf{k},\omega)]$, where $s_z$ is the spin Pauli matrix and $\sigma_0$ the sublattice identity. The resulting spectrum reproduces the graphene and spin-projected DFT bands near $\mathbf{K}$ [Fig.~\ref{fig:hbnco}(c)].
We obtain this agreement without band fitting: L\"owdin partitioning preserves the projected Kohn--Sham Green function exactly, up to the numerical errors of the finite projector shift~\cite{SM}, the frequency sampling, and the continuation.
The diagonal matrix elements $\mathcal V_{mm}(\mathbf K,\mathbf K;\omega)$ encode the exchange splitting and sublattice-asymmetric orbital shifts. Their frequency dependence comes entirely from the hybridization correction, whose resonances in Fig.~\ref{fig:hbnco}(d) identify the Co $d$ states that hybridize with graphene $p_z$ orbitals.

The sublattice asymmetry of the diagonal terms follows from the atomic positions in Fig.~\ref{fig:hbnco}(a): the $A$ carbons sit directly above the boron atoms, which occupy the fcc hollow sites of the Co surface, while the $B$ carbons sit above the centers of the hBN hexagons and the second-layer Co atoms.
We attribute the larger hybridization correction and stronger resonances of the $A$-sublattice states [Fig.~\ref{fig:hbnco}(d)] to the on-top overlap with the boron $p_z$ orbitals.

\subsection{\texorpdfstring{Graphene on PtSe$_2$: intervalley-coherent proximity coupling}{Graphene on PtSe2: intervalley-coherent proximity coupling}}

\begin{figure*}[t]
\includegraphics[width=\textwidth]{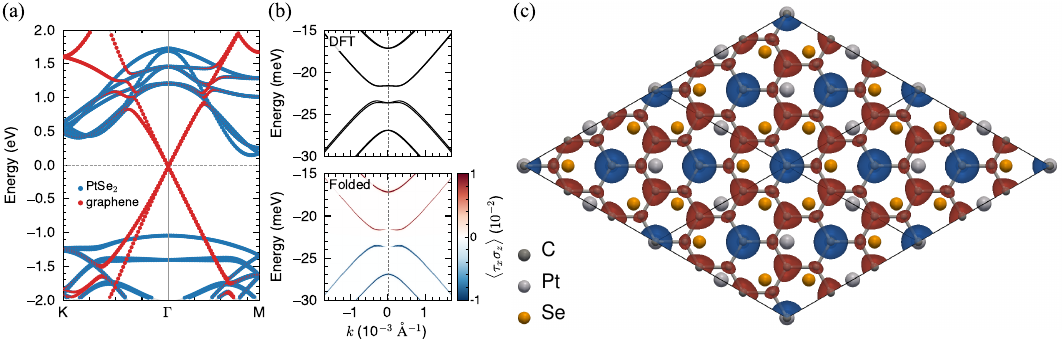}
  \caption{
    Graphene/PtSe$_2$ intervalley-coherent proximity-induced coupling.
    (a)~Layer-projected band structure along $K{-}\Gamma{-}M$, color-coded by graphene and PtSe$_2$ character.
    (b)~Zoom on the folded Dirac manifold: DFT bands (upper) and the spectral function calculated from the downfolded two-valley Hamiltonian with broadening $\eta=0.01$~meV (lower), color-coded by the spectral function projected on the intervalley operator $\tau_x\, s_0\,\sigma_z$.
    (c)~Atomic structure of the commensurate $3\times3$ graphene cell on PtSe$_2$. Isosurfaces of proximity-induced charge redistribution, color-coded by sign, reveal a $\sqrt{3}\times\sqrt{3}$ Kekul\'e modulation.
  }
  \label{fig:ptse2}
\end{figure*}
Monolayer PtSe$_2$ is a semiconducting 1T dichalcogenide with strong SOC. PtSe$_2$ can be grown epitaxially on graphene, and photoemission resolves the interlayer hybridization of the graphene $p_z$ bands~\cite{Bouaziz2025PtSe2}.
The reciprocal lattice of the zero-twist $3\times3$ graphene/$2\times2$ PtSe$_2$ cell contains $\mathbf K-\mathbf K'$~\cite{Chen2026Commensuration}. The two valleys therefore fold onto the same supercell momentum, allowing the intervalley-coherent coupling~\cite{SM}.
The layer-projected bands in Fig.~\ref{fig:ptse2}(a) place the graphene Dirac cone inside the PtSe$_2$ gap; the interfacial charge transfer shifts it below the Fermi level [Fig.~\ref{fig:ptse2}(b)].

The largest matrix elements of the intervalley proximity coupling $\mathcal V(\mathbf K,\mathbf K';\omega=0)$ at the Fermi level connect same-spin states on the graphene $A$ sublattice.
The sublattice selectivity follows from the atomic positions in the commensurate cell, which place the Pt and Se columns only below $A$ carbons.
The leading intervalley coupling is described by a Kekul\'e proximity-induced Hamiltonian,
\begin{equation}
  H_{\rm IVC} = \Delta_{\mathrm{IVC}}\,\tau_x \, s_0 \, \tfrac{1}{2}\left(\sigma_0+\sigma_z\right),
\label{eq:kekule}
\end{equation}
where $\tau_x$ is the valley Pauli matrix, $s_0$ is the spin identity, and $\tfrac{1}{2}(\sigma_0+\sigma_z)$ projects onto sublattice $A$. We absorb the Kekul\'e phase into the valley gauge and obtain $\Delta_{\mathrm{IVC}}=4.78$~meV by projecting the two-valley proximity matrix using Eq.~\eqref{eq:sm-projection}.
The intervalley coupling shares the $\sqrt{3}\times\sqrt{3}$ wavevector of the Kekul\'e bond order imaged for graphene on Cu(111)~\cite{Gutierrez2016Kekule} and of the textures that lock valley to momentum in graphene superlattices~\cite{Gamayun2018Kekule}, but is sublattice-diagonal rather than bond-centered.

Figure~\ref{fig:ptse2}(c) shows the proximity-induced charge redistribution $\Delta\rho(\mathbf r)=\rho_{\rm prox}(\mathbf r)-\rho_0(\mathbf r)$. We reconstruct $\rho_{\rm prox}$ and $\rho_0$ in the same Wannier basis from states calculated with and without the proximity operator, respectively. The intervalley term alone splits the $A$-sublattice sector by $2\Delta_{\mathrm{IVC}}=9.56$~meV; intravalley SOC and sublattice terms also contribute to the full folded spectrum [Fig.~\ref{fig:ptse2}(b)].
For plotting, we multiply the coherence spectral weight by $\eta$ and define $A_{\tau_x\sigma_z}(\mathbf{k},\omega) = -(\eta/\pi)\,{\rm Im}\,{\rm Tr}\left[(\tau_x\, s_0\,\sigma_z)\,G(\mathbf{k},\omega)\right]$. In Fig.~\ref{fig:ptse2}(b), this quantity identifies branches formed by coherent superpositions of $\mathbf K$ and $\mathbf K'$.

We retain $\mathbf K$ and $\mathbf K'$ as explicit valley quantum numbers after supercell folding and write the intravalley Hamiltonian in the spin-sublattice basis $\{A{\uparrow},\,B{\uparrow},\,A{\downarrow},\,B{\downarrow}\}$ as~\cite{Gmitra2016QSH}
\begin{equation}
\begin{split}
 H_\tau(\mathbf k)={}&\hbar v_F(\tau\sigma_xk_x+\sigma_yk_y)+\varepsilon_D\,s_0\sigma_0\\
 &+\Delta\,s_0\sigma_z+\tau s_z(\lambda_I^A\sigma_+-\lambda_I^B\sigma_-).
\end{split}
\label{eq:kp}
\end{equation}
The Hamiltonian applies near the two Dirac points, with $\mathbf k$ measured from $\mathbf K$ for $\tau=+1$ and from $\mathbf K'$ for $\tau=-1$. The Fermi velocity $v_F$ of the reference graphene Dirac cone follows from the momentum derivatives of $H_0$ at the valley. For graphene/PtSe$_2$, $\hbar v_F=5.47$~eV\,\AA, corresponding to $v_F=8.31\times10^5$~m\,s$^{-1}$. The Pauli matrices $\sigma_i$ and $s_i$ act on sublattice and spin, with identities $\sigma_0$ and $s_0$; $\sigma_\pm=(\sigma_0\pm\sigma_z)/2$ project onto sublattices $A$ and $B$. The parameters $\varepsilon_D$, $\Delta$, and $\lambda_I^{A,B}$ are the Dirac-point shift, staggered sublattice potential, and sublattice-resolved intrinsic SOC, respectively.

The intravalley projections in Eq.~\eqref{eq:sm-projection} give $\lambda_I^A=-1.04$~meV and $\lambda_I^B=+1.03$~meV. Including the reference contribution from $H_0$, the Dirac-point offset $\varepsilon_D=\tfrac14\operatorname{Tr}[H_\tau(\mathbf k=0)]$, with the trace over the four spin-sublattice states of one valley, is $-22.71$~meV, consistent with the folded Dirac levels in Fig.~\ref{fig:ptse2}(b). The effective low-energy model combines the two intravalley Hamiltonians in Eq.~\eqref{eq:kp} with the intervalley term in Eq.~\eqref{eq:kekule}. Section~S3 of the Supplemental Material specifies the individual projectors and spinor-gauge convention; Table~S1 collects the retained parameters~\cite{SM}.

\subsection{\texorpdfstring{Graphene on WSe$_2$: spatial modulation of proximity SOC}{Graphene on WSe2: spatial modulation of proximity SOC}}

\begin{figure*}[t]
\includegraphics[width=\textwidth]{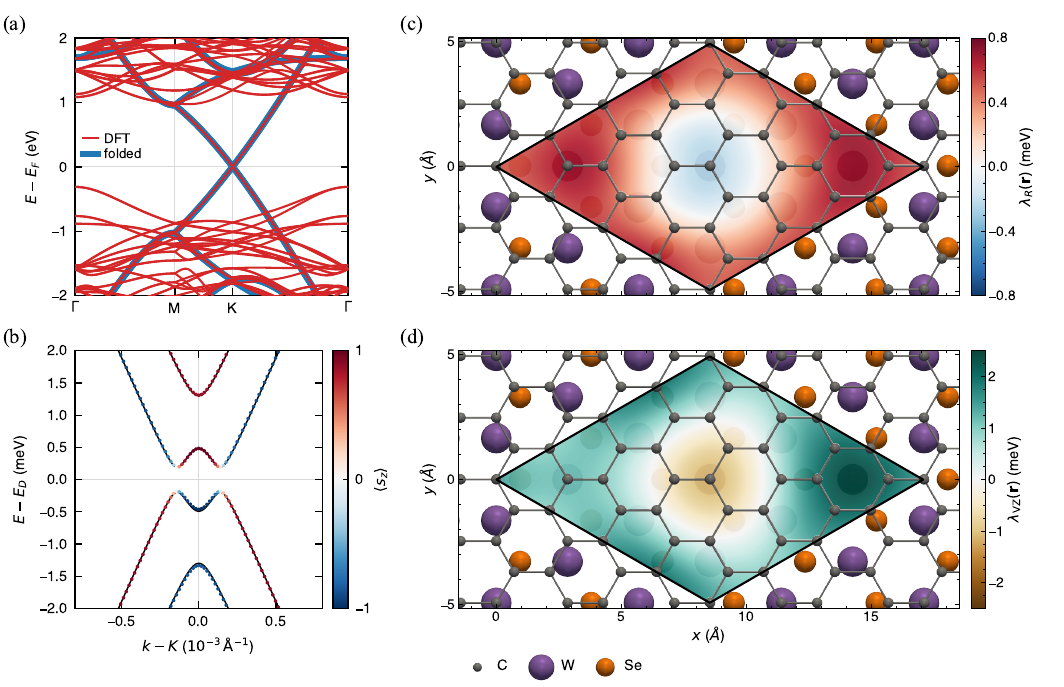}
  \caption{
    Spatially modulated proximity coupling in graphene/WSe$_2$.
    (a)~Band structure of the heterostructure supercell along $\Gamma{-}M{-}K{-}\Gamma$ (red) with the bands calculated from the downfolded Hamiltonian and folded onto the supercell (blue).
    (b)~Zoom on the Dirac point: DFT bands (black) and folded bands calculated from the downfolded Hamiltonian with the proximity coupling evaluated at the Fermi level, color-coded by the spin projection $\langle s_z\rangle$; the two sets of energies are centered independently on their respective Dirac-manifold means, so that splittings and dispersion can be compared.
    (c)~Rashba coupling $\lambda_R(\mathbf{r})$ and (d)~valley-Zeeman coupling $\lambda_{\rm VZ}(\mathbf{r})$ resolved inside the supercell, at the Fermi level.
    The black line outlines the supercell; gray, purple, and orange spheres denote C, W, and Se atoms.
  }
  \label{fig:wse2}
\end{figure*}
Lattice mismatch in graphene/WSe$_2$ creates inequivalent local environments and spatially varying proximity coupling. The nonmagnetic TMDC interface supports proximity SOC in graphene, with coupling amplitude of order $1$--$2$~meV~\cite{Wang2015, Avsar2014, Gmitra2016, Gmitra2016QSH, Garcia2018CSR, Sun2023QPI, Rao2023Ballistic}, and its strength depends on twist angle, hydrostatic pressure, and gating~\cite{Naimer2021, Fulop2021Pressure, Rockinger2026WSe2}.
We consider a zero-twist commensurate supercell of $4\times4$ graphene on $3\times3$ WSe$_2$~\cite{SM} and resolve both the uniform proximity-induced SOC and its spatial variation within the supercell.
With the proximity coupling evaluated at the Fermi level, the downfolded bands reproduce the DFT dispersion of the graphene bands after folding into the supercell Brillouin zone [Fig.~\ref{fig:wse2}(a)]. The zoom around the Dirac point in Fig.~\ref{fig:wse2}(b) resolves the meV-scale spin splittings, consistent with full heterostructure DFT calculations.

To identify the SOC contributions to the splittings, we project the momentum-diagonal proximity matrix onto spin-sublattice operators using Eq.~\eqref{eq:sm-projection}. The dominant uniform terms are the Rashba and valley-Zeeman proximity-induced couplings,
\begin{equation}
\begin{split}
 H_R&=-\lambda_R(\tau\sigma_xs_y-\sigma_ys_x),\\
 H_{\rm VZ}&=+\lambda_{\rm VZ}\,\tau\sigma_0s_z,
\end{split}
\label{eq:soc}
\end{equation}
with the spin, sublattice, and valley conventions of Eq.~\eqref{eq:kp} and $\lambda_{\rm VZ}=(\lambda_I^A-\lambda_I^B)/2$. The Rashba term mixes opposite spins, leading to the anticrossings and reversal of spin polarization in Fig.~\ref{fig:wse2}(b). The valley-Zeeman term shifts the $s_z=+1$ and $s_z=-1$ components by $+\tau\lambda_{\rm VZ}$ and $-\tau\lambda_{\rm VZ}$, respectively, so the splitting reverses between valleys. We obtain $\lambda_R=+0.35$~meV and $\lambda_{\rm VZ}=+0.80$~meV, of the same order as the values extracted from band-structure fits~\cite{Gmitra2016,Gmitra2016QSH,Naimer2021}. Section~S3 and Table~S1 of the Supplemental Material give the parameter set and the comparison between geometries~\cite{SM}.

Finite-momentum matrix elements resolve how these same couplings vary across the supercell. We use $\mathcal V_{\mathbf G}=\mathcal V(\mathbf K-\mathbf G/2,\mathbf K+\mathbf G/2;0)$, where $\mathbf G$ is a supercell reciprocal-lattice vector and the two momenta are centered on $\mathbf K$. At $\tau=+1$, the Rashba and valley-Zeeman projections and the Fourier reconstruction read
\begin{equation}
\begin{split}
 \widetilde\lambda_R(\mathbf G)&=-\frac{e^{i\mathbf G\cdot\mathbf r_{AB}}}{8}
 \operatorname{Tr}[\mathcal V_{\mathbf G}\mathcal O_R^\dagger],\\
 \widetilde\lambda_{\rm VZ}(\mathbf G)&=\frac14\big(
 e^{i\mathbf G\cdot\mathbf r_A}\operatorname{Tr}[\mathcal V_{\mathbf G}s_z\sigma_+]\\
 &\qquad+e^{i\mathbf G\cdot\mathbf r_B}\operatorname{Tr}[\mathcal V_{\mathbf G}s_z\sigma_-]\big),\\
 \lambda_c(\mathbf r)&=\sum_{\mathbf G}\lambda^{\rm int}_{c,\mathbf G}e^{i\mathbf G\cdot\mathbf r},
\end{split}
\label{eq:lambda-of-r}
\end{equation}
where $c=R,{\rm VZ}$ labels the channel, $\mathbf r$ is the in-plane position, $\mathbf r_A$ and $\mathbf r_B$ are the Wannier centers, $\mathbf r_{AB}=(\mathbf r_A+\mathbf r_B)/2$ is the bond midpoint, $\sigma_\pm$ are the sublattice projectors of Eq.~\eqref{eq:kp}, and $\mathcal O_R=\sigma_xs_y-\sigma_ys_x$ is the $\tau=+1$ Rashba operator, which enters $H_R$ in Eq.~\eqref{eq:soc} with a minus sign.
 The raw coefficients $\widetilde\lambda_c(\mathbf G)$ sample $16$ momentum transfers. We obtain the coefficients $\lambda^{\rm int}_{c,\mathbf G}$ of the interpolated fields using the reconstruction conventions~\cite{SM}.

The uniform and the first reciprocal-lattice momentum-transfer contributions dominate the spatial modulation. Each of the six shortest nonzero supercell reciprocal vectors carries a Fourier amplitude of $0.11$~meV for Rashba coupling and $0.34$~meV for valley-Zeeman coupling. Contributions from larger momentum transfers are small and can be neglected in this description; Table~S2 of the Supplemental Material lists the uniform and first-momentum-transfer Fourier coefficients.
The fields shown in Fig.~\ref{fig:wse2}(c,d) range from $-0.23$ to $+0.70$~meV for Rashba coupling and from $-1.08$ to $+2.40$~meV for valley-Zeeman coupling. Both couplings change sign within the supercell, with spatial variations comparable to their uniform components.

\section{Discussion}
We introduced a first-principles downfolding method that obtains the proximity operator $\mathcal V(\mathbf k,\mathbf k';\omega)$ in the target-layer Wannier basis without band fitting. The operator retains the energy dependence of interlayer hybridization and the momentum and spatial structure imposed by the interface. The Wannier orbitals of the isolated target layer are kept fixed, so all interface-induced symmetry breaking is contained in $\mathcal V$.

Projecting $\mathcal V$ onto spin, sublattice, and valley operators yields the coupling parameters of effective low-energy models.
The full downfolded spectra reproduce the low-energy DFT bands in all three systems [Figs.~\ref{fig:hbnco}(c), \ref{fig:ptse2}(b), and \ref{fig:wse2}(a,b)] and provide a direct check on the method. The operator supplies scattering matrix elements for $T$-matrix calculations of transport and spin relaxation. For correlated-electron calculations, $\mathcal V$ provides the proximity contribution to the self-energy, whose energy dependence comes from interlayer hybridization; electron correlations beyond the DFT starting point can be added as a separate self-energy.

The method requires a disentangled low-energy manifold whose MLWF Hamiltonian reproduces the DFT bands of the isolated target layer~\cite{Souza2001}. Future work could incorporate Hubbard~\cite{Timrov2022HP} or $GW$ quasiparticle corrections~\cite{Graml2024GW} to extend the method to a wider range of materials. The present formulation uses commensurate supercells, so the size of periodic approximants can limit the accessible twist angles. For interfaces with slowly varying local stacking, Jung et al.~\cite{Jung2014} constructed continuum models from first-principles Wannier Hamiltonians of bilayers at different relative displacements. We evaluate $\mathcal V$ in the fixed Wannier gauge of the isolated target layer, so we express matrix elements for different stackings in the same reference basis without re-wannierizing the heterostructure. Further extensions could treat superconducting interfaces and resonant defects.

\begin{acknowledgments}
We thank Jaroslav Fabian for valuable discussions.
Y.Z., J.F., and O.V.Y. acknowledge the support by the Swiss National Science Foundation (grant Nos.~204254 and 224624). First-principles calculations have been performed at the Swiss National Supercomputing Centre (CSCS) under Project No.~lp96 and the facilities of the Scientific IT and Application Support Center of EPFL.
D.K.~acknowledges National Science and Technology Council of Taiwan (No.~114-2112-M-006-034-MY3), Slovak Academy of Sciences (Project IMPULZ-2021-26 SUPERSPIN), and Slovak Research and Development Agency (contract No.~APVV-24-0134).
Y.Z.~used the AI coding assistant Claude Code (Anthropic; Claude Opus 5 and 5.5, Fable 5 and 5.1, and Sonnet 5 models) to help write and debug the calculation, analysis, and plotting scripts of this work and to revise the manuscript text, reviewed every change, and takes full responsibility for the content.
\end{acknowledgments}

\section*{Data availability}
The code, input files, and scripts behind the reported results will be made openly available upon publication.
The downfolding code used for this work, \texttt{qe-downfolding} (version 0.3.0), will be released under the MIT license at \url{https://github.com/yaroslav-zhumagulov/qe-downfolding}.
The Quantum ESPRESSO and Wannier90 input files of the three heterostructures, and the downfolding, analysis, and plotting scripts behind the reported couplings and figures will be collected in the \texttt{examples} branch of the same repository, \url{https://github.com/yaroslav-zhumagulov/qe-downfolding/tree/examples}.

\bibliography{main}

\end{document}